\documentclass[aps,prb,twocolumn,floatfix,superscriptaddress,showpacs,longbibliography]{revtex4-1}

\usepackage{graphicx}
\usepackage{amsmath}
\usepackage{amssymb}
\usepackage{soul}
\usepackage{dsfont}
\usepackage[pdftex]{color}

\newcommand{\sgn}[1]{\mathrm{sgn} \left( #1 \right)}

\def\Tr{\mathrm{Tr}}

\begin{document}
\title{Conductance fluctuations and disorder induced $\nu=0$ quantum Hall plateau in topological insulator nanowires.}
\author{Emmanouil Xypakis}
\affiliation{Max-Planck-Institut f{\"u}r Physik komplexer Systeme, 01187 Dresden, Germany}
\author{Jens H.\ Bardarson}
\affiliation{Max-Planck-Institut f{\"u}r Physik komplexer Systeme, 01187 Dresden, Germany}

\begin{abstract}
Clean topological insulators exposed to a magnetic field develop Landau levels accompanied by a 
nonzero Hall conductivity for the infinite slab geometry.
In this work we consider the case of disordered topological insulator nanowires and find, in contrast,
that a zero Hall plateau emerges within a broad energy window close to the Dirac point.
We numerically calculate the conductance and its distribution for a statistical ensemble of disordered nanowires, and use the conductance fluctuations to study the dependence of the insulating phase on system parameters,
such as the nanowire length, disorder strength and the magnetic field.
\end{abstract}
\maketitle
\section{Introduction}
That topological insulators are bulk insulators with a topologically protected metallic surface state is by now widely understood \cite{Hasan2010,Moore2010,Qi2010,Qi2011}.
Yet, when three-dimensional (3D) topological insulators are made into a nanowire, the surface state spectrum is gapped and insulating~\cite{Bardarson2013}.
The reason is fundamental: the surface Dirac fermion's spin is locked to the fermion momentum and therefore rotates by $2\pi$ when the electron goes around the nanowire circumference, inducing a $\pi$ Berry's phase \cite{Ran2008,Rosenberg2010,Ostrovsky2010,Bardarson2010}.
In the simplest case of a cylindrical wire this Berry's phase can, through a gauge transformation, be included fully via anti-periodic boundary conditions, leading to a gap that scales as $\hbar v_F/W$, with $v_F$ the fermi velocity and $W$ the wire circumference.
This gapping out of the surface states due to the Berry's phase is general, however, and does not rely on the cylindrical geometry.
The gap can be significant, despite it being in its nature a confinement gap; a $280$~nm perimeter Bi$_2$Se$_3$ or Bi$_2$Se$_3$ wire with a Fermi velocity of~$v_F = 5 \times 10^5$ m/s \cite{Chen2015,Zhang2009}, for example, has a gap of about $7$~meV.

The physical properties of such topological insulator nanowires are sensitively tuned by magnetic fields.
A magnetic flux along the length of the wire is picked up by the Dirac fermion as an Aharonov-Bohm flux when encircling the wire.
A flux $h/2e$ of a half a flux quantum exactly cancels the Berry's phase and closes the gap \cite{Ran2008,Ostrovsky2010,Bardarson2010,Rosenberg2010}. 
In the process the number of transverse modes in the wire goes from an even number (each gapped mode is doubly degenerate due to time reversal symmetry) to an odd number, with one linearly dispersing non-degenerate mode.
Even in the presence of (time-reversal preserving) disorder, there is then always at least one perfectly transmitted mode, as required by time reversal symmetry and an odd number of modes \cite{Bardarson2008}.
As a consequence, one expects to obtain Aharonov-Bohm oscillations with period $h/e$ \cite{Bardarson2010,Zhang2010} and when the odd number of modes are combined with $s$-wave superconducting proximity effect, a Majorana mode emerges at the ends of the wire \cite{Cook2011,Ilan2014}. 
The properties of topological insulator nanowires and other nanostructures, in particular the Aharonov-Bohm effect and the related gate-modulated conductance oscillations~\cite{Bardarson2010}, have been extensively studied experimentally~\cite{Kong2010,Dufouleur2013a,Peng2010a,Hong2014,Jauregui2016,Jauregui2015,Veyrat2015,Dufouleur2015}.

A magnetic field perpendicular to the length of the wire, in turn, induces chiral modes \cite{Lee2009,Vafek2011,Zhang2012,Sitte2012,Brey2014,DeJuan2014}.
This is best understood in terms of the quantum Hall effect.
The Dirac fermion at the surface sees a magnetic field whose direction is determined relative to the outward surface normal.
Since the normal rotates, one part of the surface sees a magnetic field pointing in one direction, and the other part of the surface sees a magnetic field pointing in the opposite direction.
When the magnetic field is strong enough to give rise to a quantum Hall state, therefore, one part of the surface is in the $\nu$ quantum Hall state and the other in the $-\nu$ state. 
Some part of the surface is then at the interface between two topologically distinct insulating phases and accordingly has to be gapless; exactly at these points one finds chiral modes.
In particular, since the quantum Hall states of Dirac fermions are half-integral, the difference in the topological number of the two states, which determines the number of chiral modes, is an odd integer $2n+1$ with $n\in \mathbb{Z}$ \cite{Lee2009,Sitte2012,Konig2014}.
The combination of chiral, perfectly transmitted, and Majorana modes allows for several interesting experimental setups.
For example, a normal-metal--superconductor junction in a nanowire can be used to observe the turning on and off of topological superconductivity and the emergence of a Majorana mode~\cite{DeJuan2014}, and a $p$--$n$ junction in a perpendicular magnetic field can be used to construct a spin-based Mach-Zehnder interferometer \cite{Ilan2015}.

The quantum (anomalous) Hall effect has been experimentally observed in thin films of 3D topological insulators \cite{Brune2011a,Chang2013,Xu2014,Yoshimi2014,Kou2014,Kou2015,Checkelsky2014,Xu2015,Chang2015,Feng2015}.
In these experiments, more than just the odd integer quantum Hall states were observed, and in some even a $\nu = 0$ plateau was observed \cite{Feng2015}.
The emergence of this state can be understood from the action of the back gate, which tunes the electron density on only one of the surfaces such that the bottom and top surfaces are not necessarily in the opposite quantum Hall states. 
In particular, the states of the two surfaces can be chosen such that a $\nu=0$ quantum Hall state is obtained, with zero Hall and longitudinal conductance.
The charge and spin transport properties of this $\nu=0$ state where theoretically discussed in Ref.~\onlinecite{Morimoto2015}.
A $\nu = 0$ quantum Hall state can alternatively be induced by magnetic disorder or, in graphene, by additional sublattice symmetry breaking perturbations \cite{Morimoto2015, Zhang2015,Novoselov2005,Ostrovsky2008}.

In this work we study how in topological insulator nanowires a $\nu = 0$ quantum Hall state can be induced by scalar disorder only.
In particular, around the Dirac point the two-terminal conductance goes to zero with increasing disorder strength.
Such behavior was noted in Ref.~\onlinecite{DeJuan2014} but not studied in any detail as it was not the focus of that work.
Here we numerically study the details of this conductance dip and in particular focus on the conductance fluctuations in the transition out of this state.
This work, by focusing on energies close to the Dirac point, partially complements a recent joint theoretical and experimental study of conductance fluctuations in topological insulator nanowires at high electron densities~\cite{Dufouleur2015}.

\section{Model} 
We model a 3D topological insulator nanowire with a rectangular cross section with height $H$, width $W$, and length $L$, as shown in Fig.~\ref{fig:wave_function}~i).
In this paper all data are obtained for the specific case $H=20$~nm and $W=120$~nm, unless otherwise noted. 
Assuming the bulk of the wire to be a perfect insulator, the surface state is described by the Dirac Hamiltonian \cite{Hasan2010,Moore2010,Qi2011,Bardarson2013}
\begin{equation}
	H = v_F(\mathbf{p}-e\mathbf{A})\cdot{\boldsymbol\sigma}+V(\mathbf{r}),
	\label{eq:Hamiltonian}
\end{equation}
where $v_F$ is the Fermi velocity, which we take to be equal to $v_F = 5 \times 10^5$~m/s as in Bi$_2$Te$_3$ or in Bi$_2$Se$_3$~\cite{Chen2015,Zhang2009};
the surface state momentum $\boldsymbol{p} = (p_x, p_y)$, where $x$ runs along the nanowire axis and $y \in [0,P]$ runs along the azimuthal direction with $P=2(H+W)$ the nanowire perimeter; $\boldsymbol{\sigma} = (\sigma_x, \sigma_y)$ are the Pauli matrices, and $\mathbf{A}$ is the electromagnetic vector potential that describes the magnetic field $\mathbf{B} = \mathbf{\nabla} \times \mathbf{A}$ perpendicular to the nanowire axis.
The disorder potential $V(\mathbf{r})$ is taken to be Gaussian correlated
\begin{equation}
	\langle V(\mathbf{r})V(\mathbf{r}\; ') \rangle = 
	g\frac{ (\hbar v_F)^2 }{2\xi^2}\exp\left(\frac{-|\mathbf{r}-\mathbf{r}'|^2}{2\xi^2}\right),
	\label{disorder}
\end{equation}
where $\langle \dots \rangle$ denotes the disorder average, $g$ the disorder strength and $\xi$ the disorder correlation length which we take to be $\xi = 10$~nm.
In general, the Dirac Hamiltonian~\eqref{eq:Hamiltonian} has an additional curvature induced spin-connection term that determines how the spin rotates as the Dirac fermion moves around the surface \cite{Bardarson2010,Zhang2010,Konig2014,Rosenberg2010}; here we incorporate this term through the boundary condition
\begin{equation}
	\psi (x,y+P)= -\psi(x,y).
	\label{eq:bound_cond}
\end{equation}
The appearance of the minus sign can alternatively be understood as arising from the $\pi$ Berry phase coming from the $2\pi$ rotation of the (momentum locked) spin as the Dirac fermion goes around the circumference.
While this minus sign is important in a weak magnetic field, it plays no essential role in the strong field limit we mostly work in.

The Hamiltonian~\eqref{eq:Hamiltonian} and the boundary condition~\eqref{eq:bound_cond} together define a quantum transport setup characterized by the scattering matrix
\begin{equation}
        S = \begin{pmatrix} 
        r & t^\prime \\
        t & r^\prime
        \end{pmatrix},
        \label{eq:S}
\end{equation}
which consists of the probability amplitudes of transmission $t, t^\prime$ and reflection $r,r^\prime$ by the wire.
In particular, the matrix of transmission amplitudes $t$ gives the zero temperature conductance through the Landauer formula 
\begin{equation}
G=\frac{e^{2}}{h} \Tr \;t^{\dagger} t.
\end{equation}
To obtain the scattering matrix, which relates incoming modes to outgoing modes, we follow the method introduced in Ref.~\onlinecite{Beenakker1997} to calculate the transfer matrix $T$, which relates modes in the left lead to modes in the right lead. The scattering and transfer matrices are then connected through the expression \cite{Mello1988}
\begin{eqnarray}
	T~=~\begin{pmatrix} t^{\dagger -1}&r^{\prime} t^{\prime -1} \\
		- t^{\prime -1} r&t^{\prime -1} \end{pmatrix} .
	\label{trans-scat}
\end{eqnarray}
This relation holds in the basis of modes, that is, in a basis of eigenstates normalized to carry unit current.
We model the leads with highly doped Dirac fermions and therefore the current eigenstates are the eigenstates of $\sigma_x$ \cite{Tworzydo2006}.
Let, therefore, $\varphi_n(x)  = R\int_{0}^{P}\frac{dy}{ P} e^{iq_ny}\psi(x,y) $, where the quantized azimuthal momentum
$q_n = \frac{2 \pi}{P}(n-1/2)$ is compatible with the plane wave solution in the 
azimuthal $y$ direction with anti-periodic boundary conditions and the rotation matrix
$R = (\sigma_x+\sigma_z)/\sqrt{2}$ rotates the wave function to the mode basis. 
The transfer matrix between two points $x$ and $x^\prime$, obtained by solving the Schr{\"o}dinger equation $H\psi = E\psi$ with the Dirac Hamiltonian~\eqref{eq:Hamiltonian}, is 
\begin{align}
	\varphi_n(x^{\prime})=&T_{nn^{\prime}}(x^{\prime},x)\varphi_{n^{\prime}}(x),
	\nonumber \\
	T(x^{\prime},x)=&\mathcal{P}_{x_s}  e^{\int_{x}^{x^{\prime}}dx_s\{i[E-V(x_s)]\sigma_z-iA +\sigma_x q\}}, 
	\label{transfer}
\end{align}
where $q_{nn'}= q_n \delta_{n,n'}$ and $\mathcal{P}_{x_s}$ is the path ordering operator.  
The transfer matrix~\eqref{trans-scat} is then $T = T(L,0)$.
The Fourier transformed disorder and vector potentials are given respectively by
\begin{equation}
	V_{nn^{\prime}}(x) = \int_{0}^{P} \frac{dy}{ P} e^{i(q_n - q_{n^{\prime}})y} 
V(\mathbf{r})
	\label{}
\end{equation}
and
\begin{equation}
	A_{nn'}~= \sum_{m=-M}^M e BP(-1)^{\frac{m+1}{2}}\frac{\sin (m \pi W/P)}{m^2 \pi^2}\delta_{n,n'+m},
	\label{}
\end{equation}
where M is a high-frequency cutoff that we take large enough that the conductance is independent of it.

The calculation of the transfer matrix in Eq.~\eqref{transfer} is complicated by the presence of the position ordering.
To take care of this we use the multiplicative property $T(x^\prime,x) = T(x^\prime,x^{\prime\prime})T(x^{\prime\prime},x)$ of the transfer matrix, and 
write it as a product of $N$ transfer matrices over infinitesimal intervals $dx$,
\begin{align}
	T(L,0) = & \prod_{s=1}^{N} T(x_s, x_s -dx)
	\nonumber \\
	= &\prod_{s=1}^{N} e^{dx \{i[E-V(x_s)]\sigma_z-iA +\sigma_x q\}}
	\label{trans_product}
\end{align}
where $x_s=sdx$ and $L = Ndx$.
In the second equality we have dropped the position ordering and approximated the integral with the value of the operators at the point $x_s$ multiplied with $dx$.
This approximation becomes exact when $dx \ll \xi$ the correlation length of the potential; in the actual numerical calculation we make sure to take $dx$ small enough that the conductance is independent of it.
In principle, we could now obtain the transfer matrix $T$ and from that, via Eq.\eqref{trans-scat}, the scattering matrix and thereby all linear transport properties.
The transfer matrix, however, since it relates the wave function in one lead to that in the other, has exponentially small and large eigenvalues in system size. 
The product of transfer matrices in Eq.~\eqref{trans_product} is therefore numerically unstable.
To calculate the scattering matrix we instead calculate the scattering matrix $S_{x+dx,x}$ for each slice and obtain the full scattering matrix $S_{L,0}$ as a product of all the slice scattering matrices,
\begin{equation}
	S_{L,0} = S_{dx,0} \otimes S_{2dx,dx} \otimes \dots \otimes S_{L, L-dx}.
	\label{}
\end{equation}
Here, the tensor product is obtained from the multiplicative property of the transfer matrix and its relation to the scattering matrix, and is given by
\begin{eqnarray}
        & \begin{pmatrix} 
        r_1 & t_1^\prime \\
        t_1 & r_1^\prime
        \end{pmatrix} \otimes
        \begin{pmatrix} 
        r_2 & t_2^\prime \\
        t_2 & r_2^\prime
        \end{pmatrix} =
	 \\ \nonumber
          & \begin{pmatrix} 
		 r_1+t_1^\prime r_2(1-r_1^\prime r_2)^{-1} t_1 & t_1^\prime(1-r_2 r_1^\prime)^{-1} t_2^\prime \\
		 t_2(1-r_1^\prime r_2)^{-1} t_1 & r_2^\prime+ t_2 r_1^\prime
		 (1-r_2 r_1^\prime)^{-1} t_2^\prime
        \end{pmatrix}
	\label{}
\end{eqnarray}
Alternatively, one can obtain this relation directly by considering multiple scattering from two scatterers in a row.
This process is repeated for over $1500$ disorder realizations generating a conductance distribution we use to calculate disorder averaged quantities.

We calculate the electronic density of states $\rho (E)$ by calculating the Dirac Hamiltonian~\eqref{eq:Hamiltonian} energy spectrum ${E_n}$ with antiperiodic boundary conditions for a particular disorder realization.
By approximating the Dirac delta function $\delta(E)$ by a Gaussian function $\Gamma(E)$ of a fixed width of $0.1$~meV,
we obtain the density of states for each sample,
\begin{equation}
	\rho(E) = \sum_n \delta(E-E_n) \approx \sum_n \Gamma(E-E_n).
	\label{DOS}
\end{equation}
To compute the mean density of states of a disorder sample we average over $1000$ disorder realizations.

We note that our model assumes that the bulk of the wire is a perfect insulator.
While the bulk conductivity in many experiments in not small, the advances in the material preparation of topological insulators is rapid, in particular when it comes to reducing the bulk density~\cite{Ren:2010ji,Xiong:2012fc}.
Furthermore, by not taking into account bulk effects, our model allows us to go to experimental relevant nanowire sizes and to correctly describe many essential transport properties observed in experiments~\cite{Dufouleur2015}.
\section{Results}
\begin{figure}[tb]
	\begin{center}
		\includegraphics[width=\columnwidth]{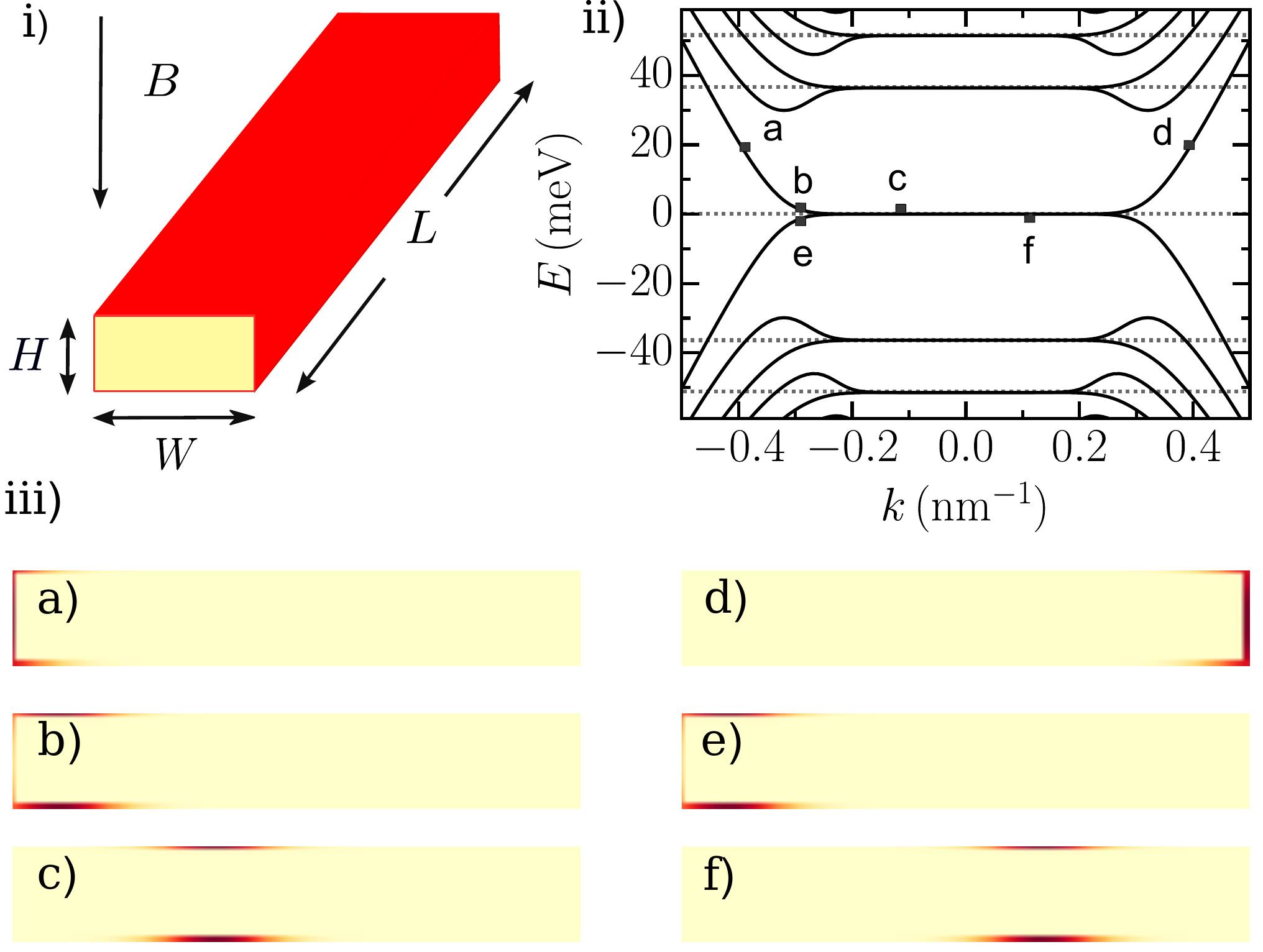}
	\end{center}
	\caption{i) A schema of a 3D strong topological insulator nanowire 
		of cross section $(H, W) =$ ($20$~nm, $120$~nm) and length $L$ with a magnetic field $B$
		perpendicular to the nanowire axis.
		ii) The energy spectrum of a clean such nanowire, exposed to a $B=4$~T magnetic field, as a function of the 
		coaxial crystal  momentum $k$. 
		iii) The probability density $|\psi|^2$ of the wave function in real space in the nanowire cross section. In all plots, the color scale goes from
		the red dark color for the maximum value to the 
		yellow light color for the minimum.
		For a state at zero energy (c, f) the wave function is
		localized both at the top and at the bottom of the wire, 
		while for higher momenta $k = (-0.4, -0.3, 0.4, -0.3)$, corresponding to (a, b, d, e), it is localized at the edges revealing its chiral nature.
		The dashed line represents the theoretical values of the relativistic Landau levels.
	}
	\label{fig:wave_function}
\end{figure}
We start by discussing the electronic structure of a nanowire in a magnetic field in the absence of disorder.
In Fig.~\ref{fig:wave_function}~ii) we plot the band structure as a function of the momentum $k$ along the length of the wire, for a magnetic field of $B=4$~T.
For small momenta $k$ we obtain flat bands that correspond to Landau levels in the top and bottom surfaces; the corresponding probability density $|\psi|^2$ shown in panel iii), for the points c and f marked in panel ii), is localized in the top and bottom surfaces.
The energies of these bulk Landau levels are consistent with that of the half integer quantum Hall 
effect of Dirac fermions~\cite{Aharonov1979,Jackiw1984,Konig2014}: $E_n = \sgn{n} \sqrt{2eB |n|} v_F$.
The Landau level center shifts with increasing momentum towards the side surfaces and eventually the modes become dispersive when they reside on the edge surfaces---see the plot of the wave functions a, b, d, and e in panel iii).
One can understand this in terms of the quantum Hall effect and the bulk-boundary correspondence, with one important difference:
the edge surfaces are not a physical edge in the true sense, but rather are the interface between two distinct quantum Hall states, one at the top surface and one at the bottom surface.
Since from the perspective of the surface state Dirac fermion, the magnetic field points in the opposite direction, these two states have opposite Landau level index.
The half-integer Dirac fermion Landau levels then mean that the number of edge 
states is odd, and at low energies, in particular, only a single chiral edge mode is obtained \cite{Lee2009,DeJuan2014}

\begin{figure}[tb]
	\begin{center}
		\includegraphics[width=\columnwidth]{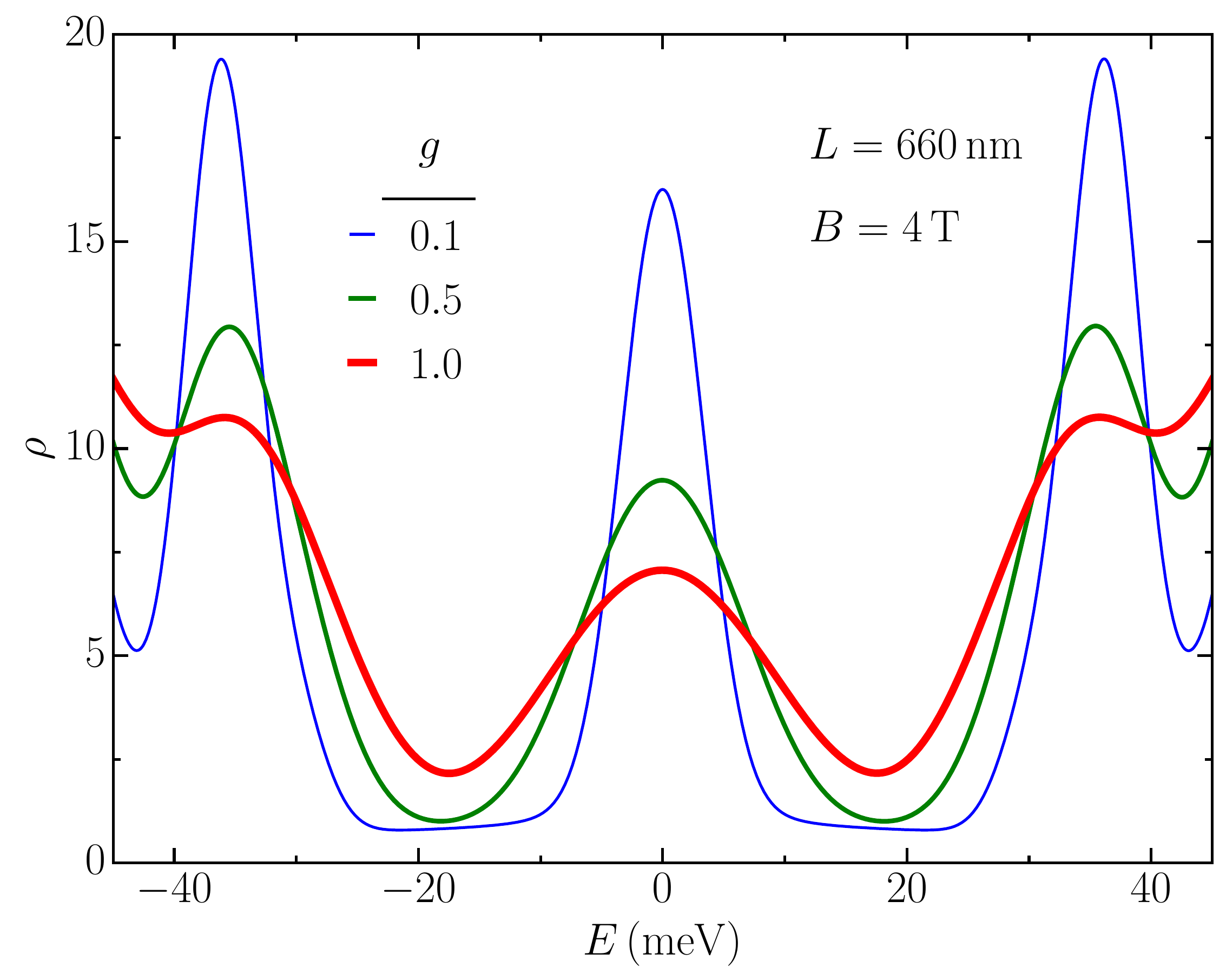}
	\end{center}
	\caption{The density of state $\rho$ as a function of the energy $E$ in a disordered, $660$nm
		long nanowire exposed to a 4 T magnetic field for three different disorder strengths $g$.
	}
	\label{fig:DoS}
\end{figure}
\begin{figure}[tb]
	\begin{center}
		\includegraphics[width=\columnwidth]{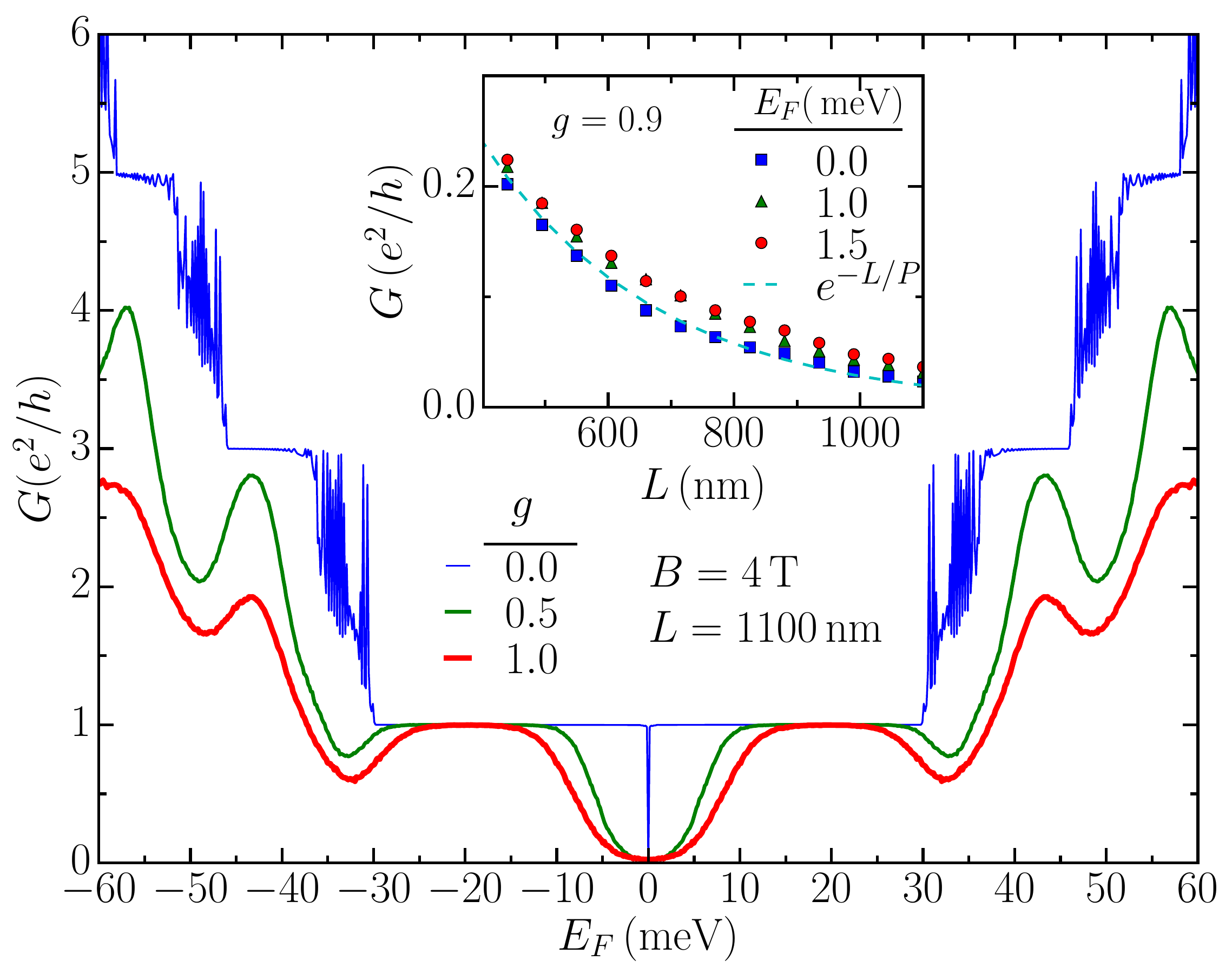}
	\end{center}
	\caption{Average conductance $G$ at disorder strength $g$ as a function of the 
		chemical potential $E_F$ for a $1100$~nm long nanowire in a $4$~T magnetic field.
		 Inset: Length dependence of the conductance for fixed disorder strength $g=0.9$ at three different chemical potentials $E_F$. 
		 The dashed line represent an exponential fit with the decay length equal to the nanowire perimeter $P$.
	}
	\label{fig:g_inset}
\end{figure}
In the presence of disorder $V$ the Landau levels broaden, as shown in Fig.~\ref{fig:DoS}.
The curves in this figure are for a disordered sample of a fixed magnetic field $B=4$~T and length $L=660$~nm, for three different disorder strengths $g$. 
Analogously to the integer quantum Hall effect in a two-dimensional electron gas, we expect that all the states are spatially localized, except the extended bulk states at the topological phase transition, i.e., when the chemical potential is tuned exactly to a Landau level.
\begin{figure}[tb]
	\begin{center}
		\includegraphics[width=\columnwidth]{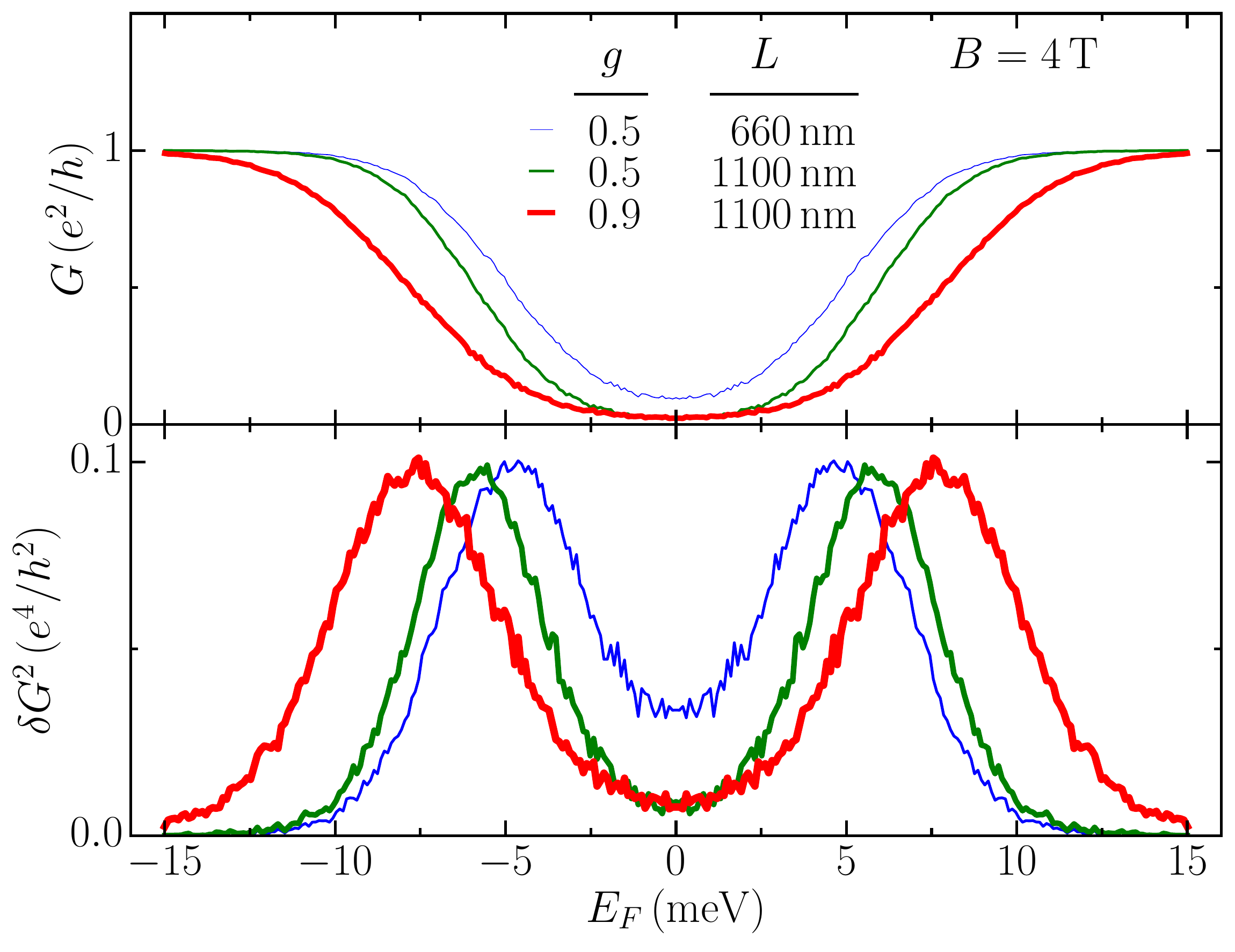}
	\end{center}
	\caption{Average conductance $G$ and the corresponding conductance
		variance $\delta G^2$ as a function of the 
		chemical potential $E_{F}$ for nanowires of various lengths $L$  and disorder strengths $g$ in a $4$~T magnetic field.
		}
	\label{fig:GL}
\end{figure}

As we show in Fig.~\ref{fig:g_inset}, for a clean topological insulator nanowire the conductance is quantized to odd integer values when the chemical potential is tuned between Landau levels and it jumps from $-e^{2}/h$ to $e^{2}/h$ being zero only at the Dirac point.
Moreover, Fabry-P{\'e}rot oscillations from first to second plateau are manifest because of intermode scattering.
The higher $\nu=2$ and $\nu=3$ quantum Hall plateaus are still visible, though their energy range becomes smaller and the quantization less good with higher chemical potentials. 
This is the topological insulator nanowire version of the half-integer quantum Hall effect.

The introduction of disorder has several qualitative effects on the conductance.
First of all, it smears out all the sharp Fabry-P{\'e}rot resonances and for these parameters, the higher Landau levels overlap in such a way that there are no clear plateaus.
%
%
In fact, only the $g_{xy} = \pm 1/2 $ states are robust to disorder in this parameter range. 
This mainly occurs because the Landau level spacing becomes smaller at higher energies and comparable to the level disorder broadening as shown in Fig. 2.
As a result, lower filling factor chiral states can scatter with higher filling factor states around the spectrum pockets of Fig.~\ref{fig:wave_function}. 
Therefore, the conductance is nonmonotonic and even drops below $1$ when the higher Landau levels start contributing (for example around 35 meV).
The $\nu=\pm 1$ Hall plateaus, however, persist over a significant energy range, and with increasing disorder and wire length a $\nu=0$ Hall plateau emerges.

In the following we focus our attention on this disorder induced zero Hall plateau.
In Fig.~\ref{fig:GL}, we plot the average conductance $G$ for a disordered nanowire ensemble and show that, when the wire length is sufficiently large and the disorder is strong, a $\nu=0$ Hall plateau is present.
The conductance drops exponentially to zero for a broad energy window close to zero energy as we show in the inset of Fig.~\ref{fig:g_inset}.
The energy regime where the $\nu=0$ Hall plateau is present, is proportional to the zero Landau level energy broadening and to the nanowire length.
For slightly higher energies, however, the conductance is quantized to one conductance quantum retaining the quantum Hall plateau.
To understand the chiral nature of the transport modes in the presence of disorder, we focus on the first quantum Hall plateau that is robust for a broad energy window.

\begin{figure}[h]
	\begin{center}
		\includegraphics[width=\columnwidth]{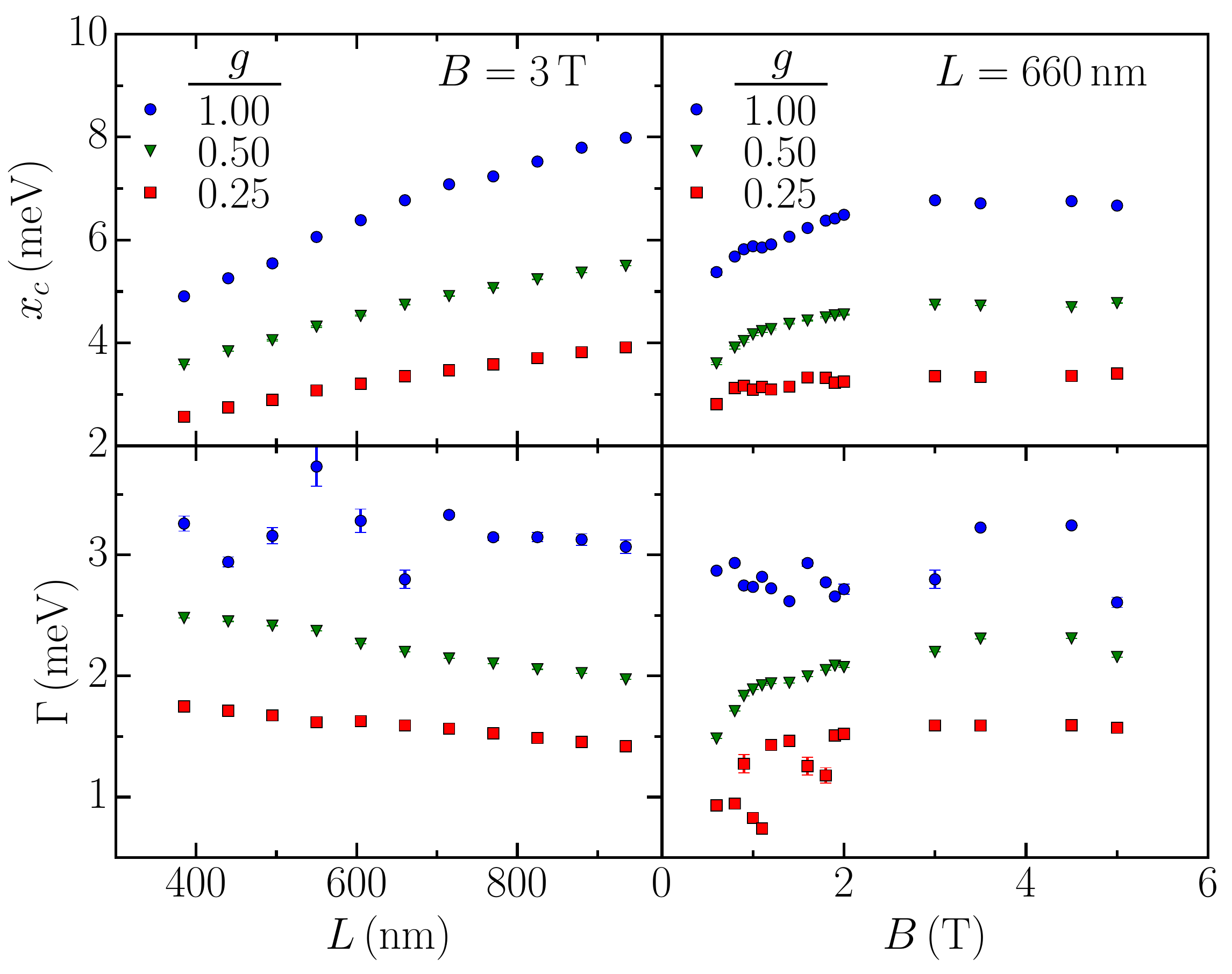}
	\end{center}
	\caption{Dependence of the conductance variance peak center $x_c$ and width $\Gamma$, obtained from a Gaussian fit,
		on the nanowire length $L$ and magnetic field $B$ for 
		three different disorder strengths $g$.
		}
	\label{fig:x_cBL}
\end{figure}

\begin{figure}[tb]
	\begin{center}
		\includegraphics[width=\columnwidth]{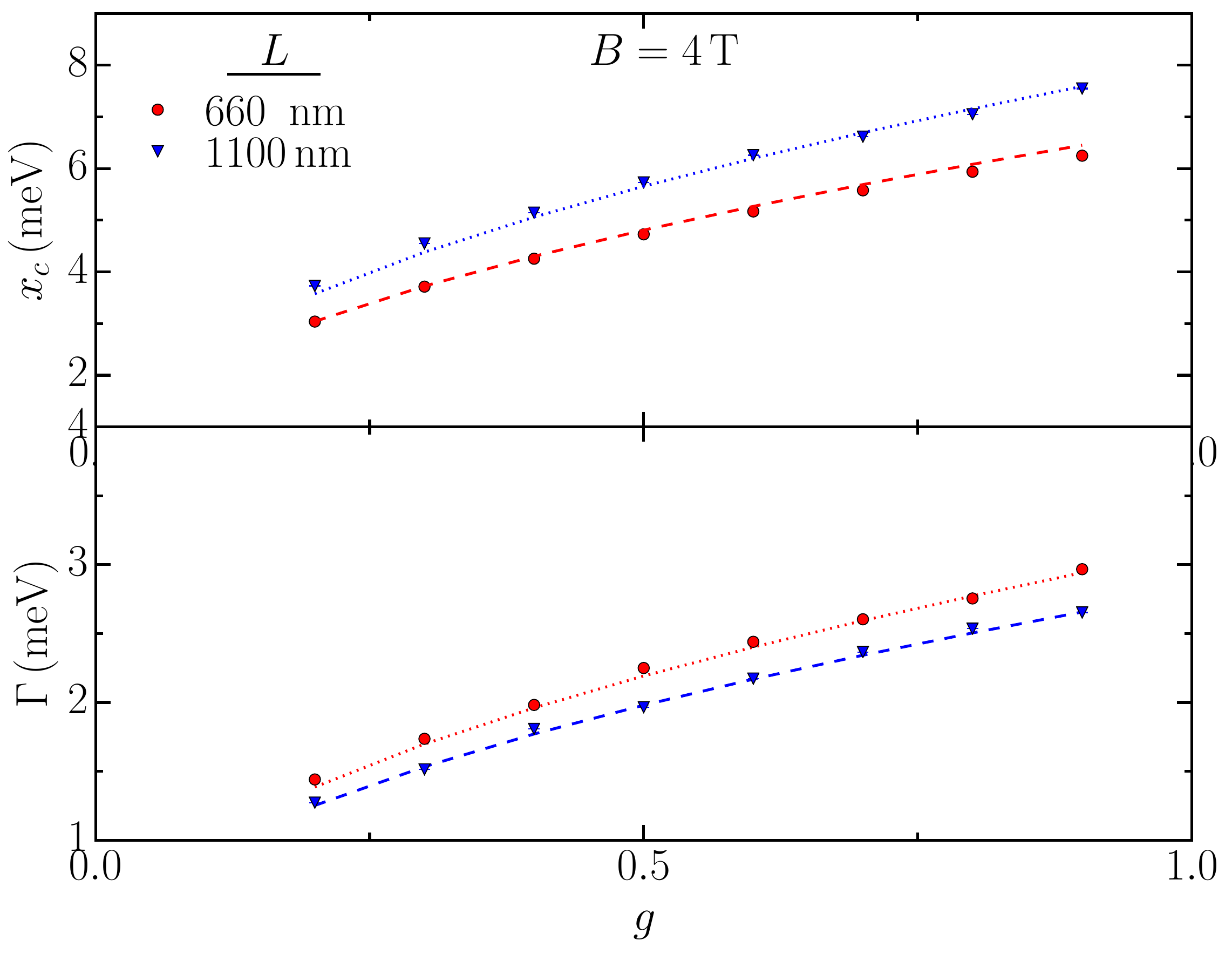}
	\end{center}
	\caption{The conductance variance peak center
		$x_c$ and width $\Gamma$ as a function of the disorder strength $g$
		for two different nanowire lengths $L$ and
		$4$~T magnetic field. The dashed lines in the plots are 
		$\sqrt{g}$ fits.
		}
	\label{fig:x_cgi}
\end{figure}

As a measure of where the chiral modes start being robust, we study the conductance variance.
In Fig.~\ref{fig:GL} we calculate the conductance variance, and show that
when the system is either in the chiral regime or does not conduct the variance is minimized, but in between there is a peak indicating the crossover.
To explore the low energy crossover we extract the center $x_c$ and width $\Gamma$ of the conductance fluctuation peak by locally approximating it with a Gaussian function. 
In Fig.~\ref{fig:x_cBL} we show that larger wires tend to exhibit a non chiral state for larger energy windows, since the conductance variance peak $x_c$ scales almost linearly with the nanowire length $L$.
This is expected since in the limit $L\rightarrow\infty$, keeping all other parameters fixed, we go to the 1D limit, in which there is no quantum Hall effect.
In this context, we note that even the long $L\gtrsim 1$ $\mu$m nanowires exhibit clear signatures of the 2D quantum Hall effect.
In addition, where the chiral modes are robust is weakly correlated to the magnetic field B.
The crossover sharpness is also uncorrelated with both the wire length $L$ and the  magnetic field $B$, as the subplots for the variance width $\Gamma$ indicate.
This demonstrates the fact that the zero Landau level is induced by the disorder and is absent in the absence of disorder.
In Fig.~\ref{fig:x_cgi} we show how $x_c$ and $\Gamma$ depend on the disorder strength $g$.
Both are correlated to the disorder strength and are almost proportional to $\sqrt{g}$.
The later fact confirms that the non chiral state energy window is proportional to the disorder broadening Eq.~\eqref{disorder}.

\begin{figure}[h]
	\begin{center}
		\includegraphics[width=\columnwidth]{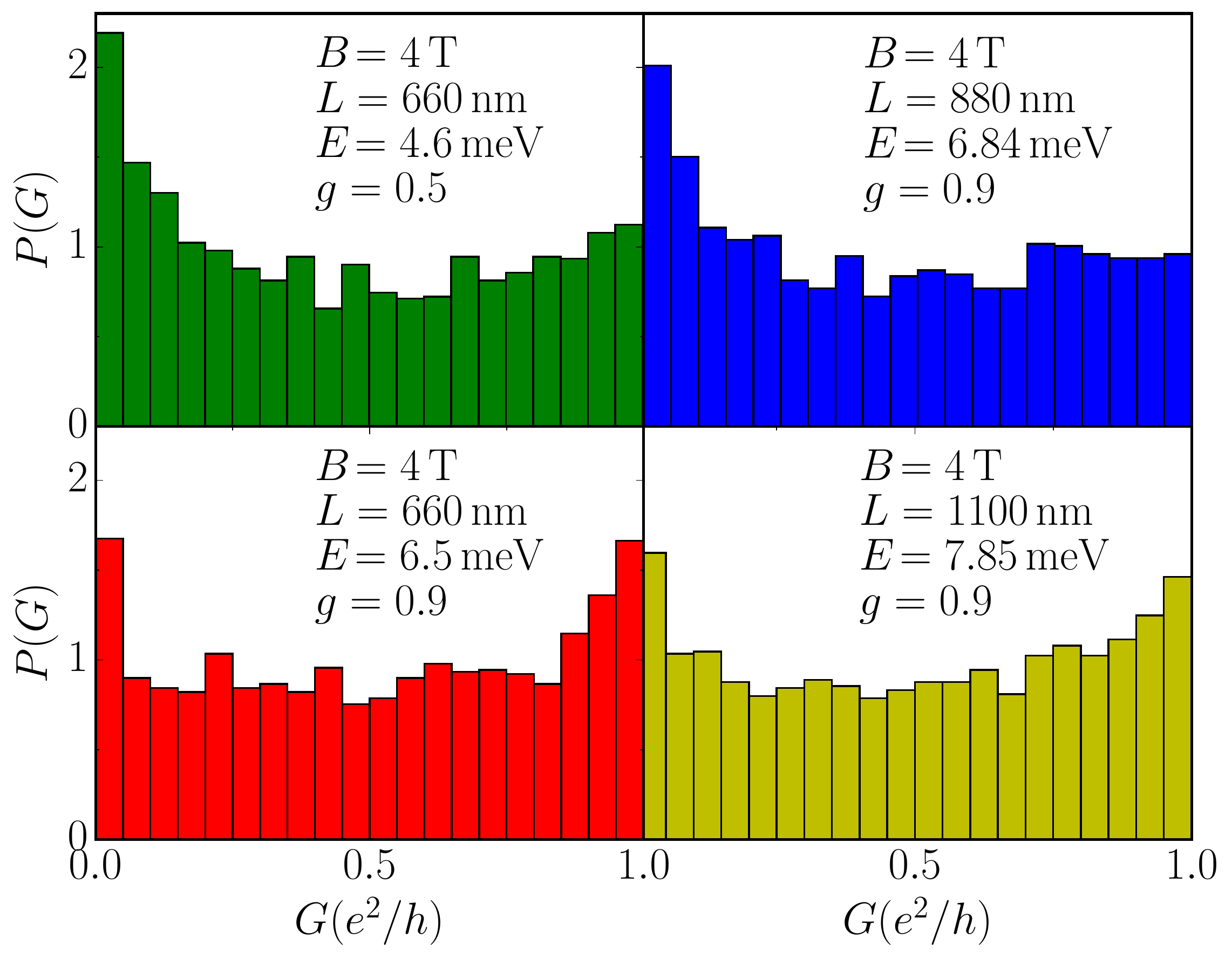}
	\end{center}
	\caption{The conductance distribution at the 
		variance peak point $x_c$
		for various disordered nanowire statistical ensemble of
		a $4$~T magnetic field but for different length $L$, disorder strength
		$g$, and energy $E$.}
	\label{fig:histo}
\end{figure}

Moreover, in Fig.~\ref{fig:histo}, we show the conductance distribution at the crossover point for different lengths and disorder
strengths.
This distribution is approximately uniform, with peaks at zero and one conductance quantum.
This is consistent with what is observed in the quantum Hall plateau transition in two-dimensional electron systems~\cite{Cho1997a,Evers2008}.

\begin{figure}[b]
	\begin{center}
		\includegraphics[width=\columnwidth]{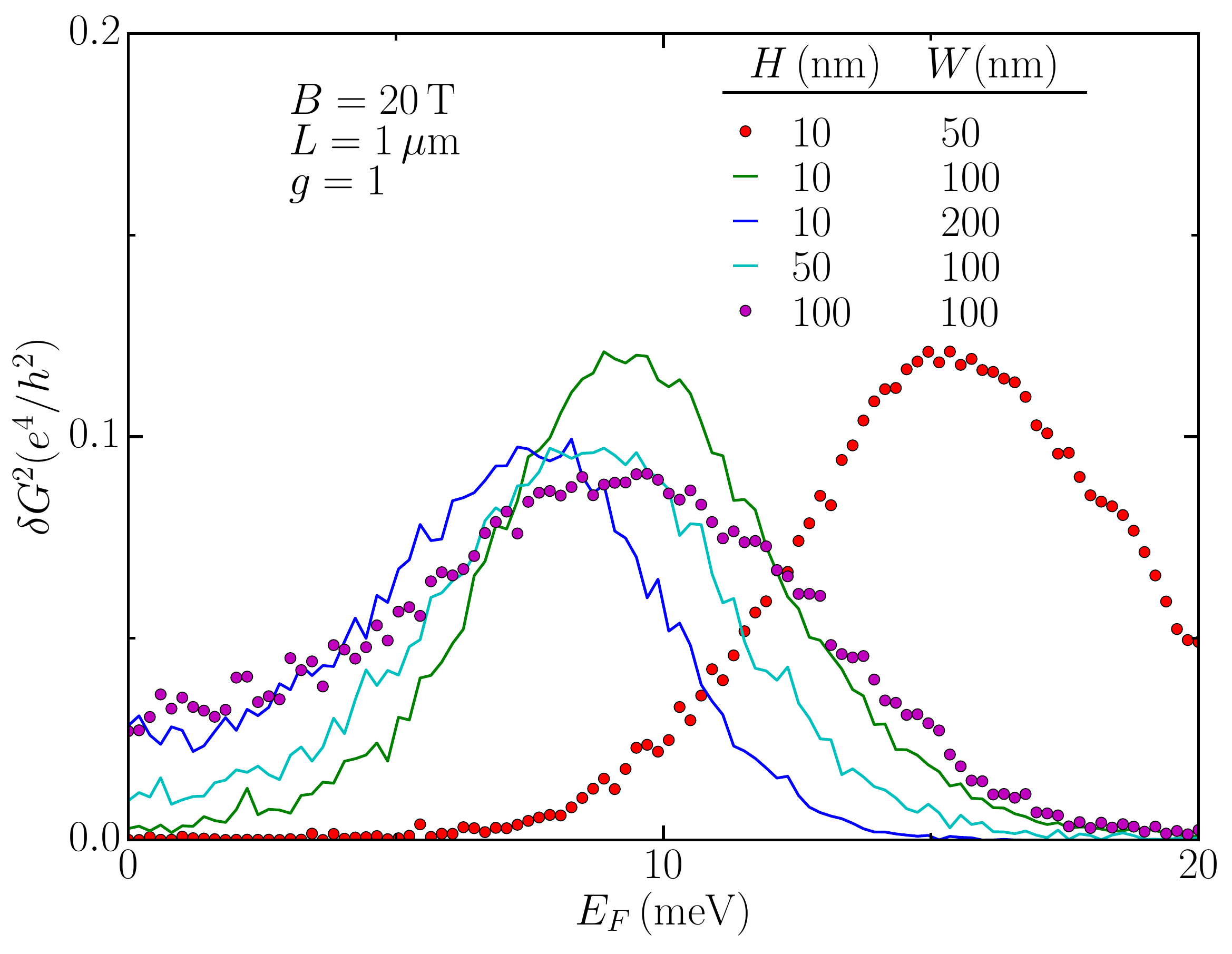}
	\end{center}
	\caption{Conductance variance $\delta G ^2$ for fixed disorder strength $g = 1.0$ and fixed length $L = 1000$~nm and a magnetic field $B = 20 $~T for various heights $H$ and widths $W$.
}
	\label{fig:perimeter}
\end{figure}

The physical mechanism for the emergence of the zero quantum Hall plateau is that counterpropagating modes on opposite side edges can couple and localize each other when their localization length $\lambda(E_F, g)$, which in the limit of infinite geometry peaks at zero energy, is comparable to the nanowire width $W$.
This is consistent with the expectation that in the limit of large $L$, with all other parameters kept fixed, we are in the 1D limit in which there is no quantum Hall effect.
The observed transition is therefore strictly a finite size crossover effect, as demonstrated in Fig.~\ref{fig:perimeter}.
By increasing the nanowire width $W$, at fixed nanowire length $L$ and height $H$, we approach the two dimensional limit, in which no $\nu=0$ quantum Hall plateau is present.

%
%
\section{Summary and discussion}

In this work we studied the transport properties of three-dimensional disordered topological insulator nanowires in a magnetic field perpendicular to the long axis.
We showed that the system exhibits a version of the half integer quantum Hall effect with one important difference. 
When the chemical potential is tuned close to the Dirac point and disorder is strong, a zero conductance Hall plateau emerges.
This is due to coupling by disorder between the Landau levels in the side surfaces of the nanowire.
This is similar to the mechanism of valley coupling graphene, which can also induce integer Hall plateaus~\cite{Ostrovsky2008}, with the important difference that there the coupling happens throughout the 2D material.
In our numerical analysis we used experimentally relevant parameters demonstrating that our results and predictions are relevant to both current and future experiments. 
For the long term goal of experimentally observing signatures of Majorana fermions on topological insulator nanowires, it is important to be able to characterize the normal state of the wire in great detail.
The conductance fluctuations studied here provide one more important and accessible quantity to study for that aim.
\section*{ACKNOWLEDGEMENT}
This work was supported by the ERC Starting Grant QUANTMATT No. 679722.
\bibliography{reference.bib}
\end{document}